\documentclass{elsart}

\usepackage{graphicx}

\begin{document}

\begin{frontmatter}

\title{On exact and perturbation solutions to nonlinear equations for heat transfer
models}
\author{Francisco M. Fern\'{a}ndez\thanksref{FMF}}

\address{INIFTA (UNLP, CCT La Plata-CONICET), Divisi\'{o}n Qu\'{i}mica Te\'{o}rica,\\
Blvd. 113 y 64 (S/N), Sucursal 4, Casilla de Correo 16,\\
1900 La Plata, Argentina}

\thanks[FMF]{e--mail: fernande@quimica.unlp.edu.ar}

\begin{abstract}
We analyze some exact and approximate solutions to nonlinear equations for
heat transfer models. We prove that recent results derived from a method
based on Lie algebras are either trivial or wrong. We test a simple
analytical expression based on the hypervirial theorem and also discuss earlier
perturbation results.
\end{abstract}

\end{frontmatter}

\section{Introduction}

In a recent paper Moitsheki et al\cite{MHM09} argued that a method based on
Lie algebras is suitable for obtaining the solution to nonlinear ordinary
differential equations that appear in simple models for heat transfer. They
compared the analytical solutions with other results coming from
perturbation approaches like homotopy perturbation method (HPM) and homotopy
analysis method\ (HAM)\cite{A06c,G06,RGT07,SH08}. It is worth noticing that
there is an unending controversy between the users of those fashionable
perturbation approaches that arose some time ago\cite{H04,L05}.

The purpose of this paper is to determine the usefulness of the results for
the heat transfer systems provided by the Lie algebraic method and those
perturbation approaches. In Sec.~\ref{sec:exact} we analyze the exact
solutions arising from Lie algebras, in Sec.~\ref{sec:Taylor} we outline
the application of the well known Taylor--series approach,
in Sec.~\ref{sec:Virial} we derive a simple accurate analytical
expressions for one of the models and in Sec.~\ref{sec:conclusions} we
summarize our results and draw conclusions.

\section{Exact solutions}

\label{sec:exact}

The first example is the nonlinear ordinary differential equation\cite{MHM09}
\begin{eqnarray}
\lbrack 1+\epsilon u(x)]u^{\prime \prime }(x)+\epsilon u^{\prime }(x)^{2}
&=&0  \nonumber \\
u(0)=1,\;u(1) &=&0  \label{eq:ex_1}
\end{eqnarray}
where the prime denotes differentiation with respect to the variable $x$.
This equation is trivial if one rewrites it in the following way $%
[(1+\epsilon u)u^{\prime }]^{\prime }=0$\cite{RGT07} and the solution is
\begin{equation}
u(x)=\frac{\sqrt{(1+\epsilon )^{2}+[1-(1+\epsilon )^{2}]x}-1}{\epsilon }
\label{eq:u_ex_1}
\end{equation}
Moitsheki et al\cite{MHM09} derived exactly this result by means of a rather
lengthy algebraic procedure. It is clear that in this case the Lie algebraic
method gives us the same answer that we can obtain in a simpler way.

For the second example
\begin{eqnarray}
u^{\prime \prime }(x)-\epsilon u(x)^{4} &=&0  \nonumber \\
u^{\prime }(0)=0,\;u(1) &=&1  \label{eq:ex_2}
\end{eqnarray}
the authors derived the simple analytical expression\cite{MHM09}
\begin{equation}
u(x)=\left( \sqrt{\frac{9\epsilon }{10}}x+1-\sqrt{\frac{9\epsilon }{10}}%
\right) ^{-2/3}  \label{eq:u_ex_2}
\end{equation}
They argued correctly that it satisfies $u(1)=1$ but they were wrong when
they stated that ``However, $u^{\prime }(0)=0$ only if $\epsilon =10/9$''.
Notice that the function $u(x)=x^{-2/3}$ that comes from such value of $%
\epsilon $ does not have the correct behaviour at $x=0$. Therefore, in this
case the Lie algebraic approach led to a wrong result.

Other authors have applied HPM and HAM to the equation\cite{A06c,G06}
\begin{eqnarray}
\lbrack 1+\epsilon u(x)]u^{\prime }(x)+u(x) &=&0  \nonumber \\
u(0) &=&1  \label{eq:ex_3}
\end{eqnarray}
with the trivial solution
\begin{equation}
\ln u(x)+\epsilon [u(x)-1]+x=0  \label{eq:u_ex_3}
\end{equation}

In the following two sections we discuss some of these problems from
different points of view.

\section{Taylor series}

\label{sec:Taylor}

If the variable of the nonlinear equation is restricted to a finite
interval, one can try a straightforward power--series solution $%
u(x)=u_{0}+u_{1}x+u_{2}x^{2}+\ldots $ and obtain the unknown model parameter
from the boundary conditions. In the case of the example (\ref{eq:u_ex_1})
the radius of convergence of this series is $(\epsilon +1)^{2}/[\epsilon
(\epsilon +2)]$ and therefore the approach will be useful for small and
moderate values of $\epsilon $. As $\epsilon $ increases the rate of
convergence of the Taylor--series method decreases because the radius of
convergence approaches unity from above. However, this example is trivial
and of no interest whatsoever for the application of a numerical or
analytical method. This reasoning also applies to example (\ref{eq:ex_3})
although in this case we do not have an explicit solution $u(x)$ but $x(u)$.

The example (\ref{eq:ex_2}) is more interesting because there appears to be
no exact solution, and for this reason we discuss it here. The unknown
parameter is $u(0)=u_{0}$ and the partial sums for the Taylor series about $%
x=0$%
\begin{equation}
u^{[N]}(x)=\sum_{j=0}^{N}u_{j}(u_{0})x^{j}  \label{eq:u_x_series}
\end{equation}
enable one to obtain increasingly accurate estimates $u_{0}^{[N]}$ as $N$
increases. Such estimates are roots of $u^{[N]}(1)=1$. Although the rate of
convergence decreases as $\epsilon $ increases it is sufficiently great for
most practical purposes. Notice that the HAM perturbation corrections for
this model are polynomial functions of $x$\cite{A06c} whereas the HPM has
given polynomial functions of either $x$\cite{G06} or $e^{-x}$\cite{RGT07}.
However, there is no doubt that the straightforward power--series approach
is simpler and does not require fiddling with adjustable parameters\cite
{A06c,SH08}.

\section{The hypervirial theorem}

\label{sec:Virial}

The analysis of the nontrivial equations for heat transfer models may be
easier if we have simple approximate analytical solutions instead of
accurate numerical results or cumbersome perturbation expressions. In the
case of the models (\ref{eq:ex_1}) and (\ref{eq:ex_3}) there is no doubt
that the exact analytical expressions should be preferred. For that reason,
in what follows we concentrate on the seemingly nontrivial model (\ref
{eq:ex_2}).

We have recently shown that the well known virial theorem may provide simple
analytical solutions for some nonlinear problems\cite{AF09a,AF09b}. In
particular, we mention the analysis of a bifurcation problem that appears in
simple models for combustion\cite{AF09a}. The only nontrivial problem
outlined above is a particular case of nonlinear ordinary differential
equations of the form
\begin{eqnarray}
u^{\prime \prime }(x) &=&f(u(x))  \nonumber \\
0 &\leq &x\leq 1  \label{eq:gen_nonlin}
\end{eqnarray}

The hypervirial theorem is a generalization of the virial one. If $w(u)$ is
an arbitrary differentiable weight function, the hypervirial theorem
provides the following suitable expression for our problem (\ref
{eq:gen_nonlin}):
\begin{eqnarray}
\int_{0}^{1}[w(u)u^{\prime }]^{\prime }dx &=&w(u(1))u^{\prime
}(1)-w(u(0))u^{\prime }(0)  \nonumber \\
&=&\int_{0}^{1}\left[ \frac{dw}{du}(u^{\prime })^{2}+w(u)f(u)\right] dx
\label{eq:VT_gen}
\end{eqnarray}
In the particular case of the example (\ref{eq:ex_2}) we have
\begin{equation}
w(1)u^{\prime }(1)=\int_{0}^{1}\left[ \frac{dw}{du}(u^{\prime
})^{2}+\epsilon w(u)u^{4}\right] dx  \label{eq:VT_ex_2}
\end{equation}
When $w(u)=u$ we obtain the virial theorem. Here we also consider the even
simpler choice $w(u)=1$ that we will call hypervirial although it is just
a particular case.

Since $u^{\prime \prime }(x)>0$ we try the ansatz
\begin{equation}
u_{app}(x)=\frac{\cosh (bx)}{\cosh (b)}  \label{eq:u_app}
\end{equation}
that satisfies the boundary conditions in equation (\ref{eq:ex_2}). It
follows from equation (\ref{eq:VT_ex_2}) that the adjustable parameter $b$
is a root of
\begin{eqnarray}
&&3e^{10b}(5b^{2}-2\epsilon )+5e^{8b}(12b^{3}+9b^{2}-10\epsilon
)+30e^{6b}(6b^{3}+b^{2}-10\epsilon )  \nonumber \\
&&+30e^{4b}(6b^{3}-b^{2}+10\epsilon )+5e^{2b}(12b^{3}-9b^{2}+10\epsilon )
\nonumber \\
&&-3(5b^{2}-2\epsilon )=0  \label{eq:b(epsilon)1}
\end{eqnarray}
when $w(u)=u$ and
\begin{eqnarray}
&&3e^{10b}(5b^{2}-\epsilon )+5e^{8b}(9b^{2}-5\epsilon
)+30e^{6b}(b^{2}-5\epsilon )+30e^{4b}(5\epsilon -b^{2})  \nonumber \\
&&+5e^{2b}(5\epsilon -9b^{2})-3(5b^{2}-\epsilon )=0  \label{eq:b(epsilon)0}
\end{eqnarray}
when $w(u)=1$.

\begin{figure}[H]
\begin{center}
\includegraphics[width=9cm]{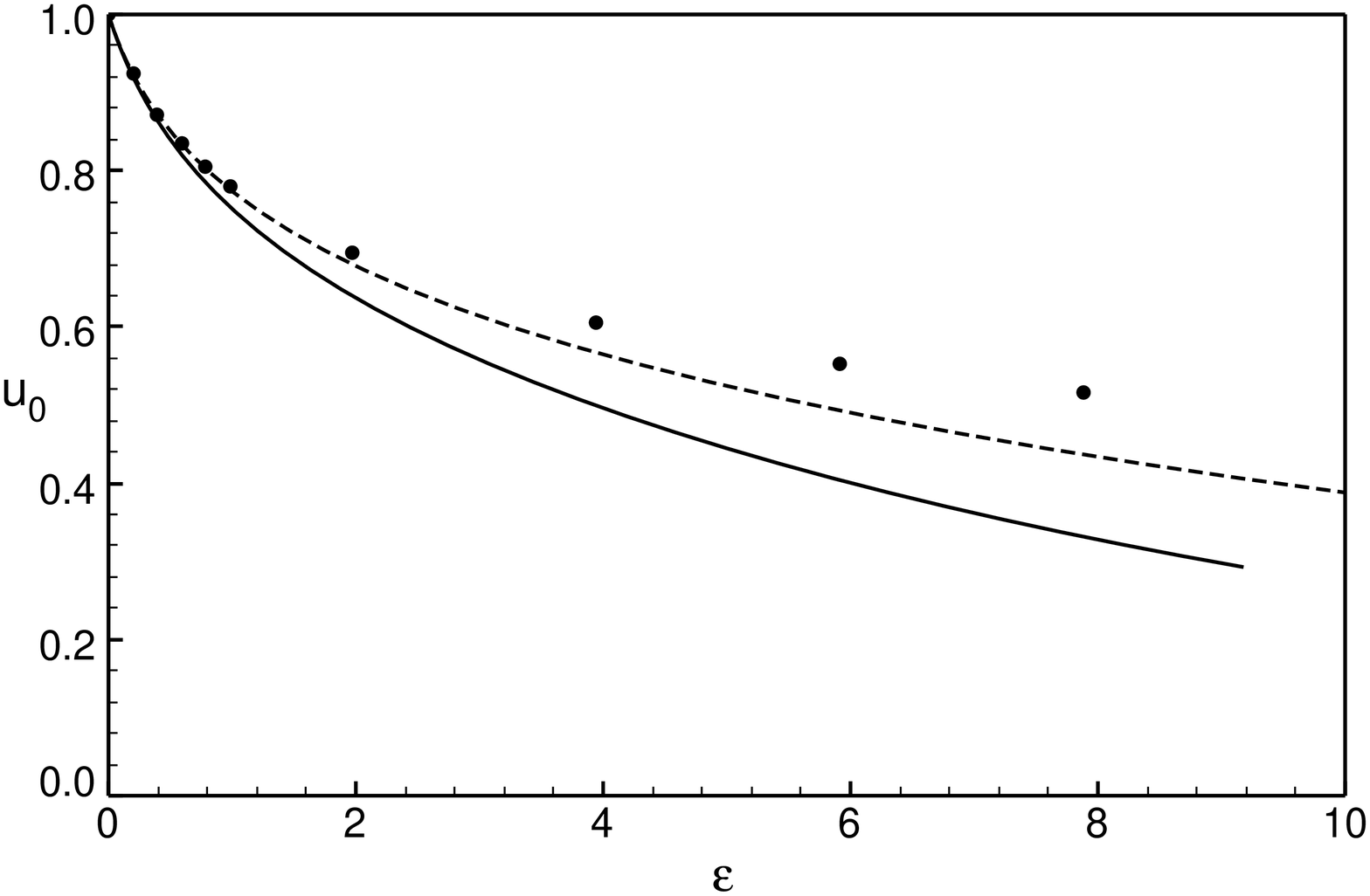}
\end{center}
\caption{Numerical (circles) virial (solid line) and hypervirial (dashed
line) values of $u_0$.}
\label{fig:HT1}
\end{figure}

Fig.~\ref{fig:HT1} shows $u_{app}(0)$ for some values of $\epsilon $ and
also the accurate result obtained from the Taylor series discussed in Sec.~%
\ref{sec:Taylor}. We appreciate that the accuracy of the analytical
expression (\ref{eq:u_app}) decreases as $\epsilon $ increases. However, if
one takes into account the simplicity of equation (\ref{eq:u_app}) the
agreement is remarkable. Besides, the hypervirial theorem with $w=1$ proves
to be more accurate than the virial theorem. It is curious that there is no
such test for the HPM or HAM\cite{A06c,G06}.

As a particular example we consider $\epsilon =0.7$ (the preferred parameter
value for both HAM and HPM calculations\cite{A06c,G06}). From the partial
sums of the Taylor--series with $N\leq 30$ we obtain $u_{0}=0.8186424785$.
The analytical function (\ref{eq:u_app}) yields $b\approx 0.70$,
$u_{app}(0)\approx 0.80$ for $w=u$ and $b=0.657$, $u_{app}(0)\approx 0.817$
for $w=1$ that is a reasonable estimate of the unknown parameter. Again we see that
the hypervirial approach is better than the virial one. Fig.~\ref{fig:HT2}
shows accurate values of $u(x)$ given by the Taylor series with $N=30$, our
approximate analytical virial expression $u_{app}(x)$ and equation (\ref
{eq:u_ex_2}) for $0\leq x\leq 1$. It seems that the accuracy of $u_{app}(x)$
is somewhat between the HAM results of 5th and 10th order\cite{A06c}. On the
other hand, the equation (\ref{eq:u_ex_2}) derived by the Lie algebraic
method\cite{MHM09} exhibits a wrong behaviour.

\begin{figure}[H]
\begin{center}
\includegraphics[width=9cm]{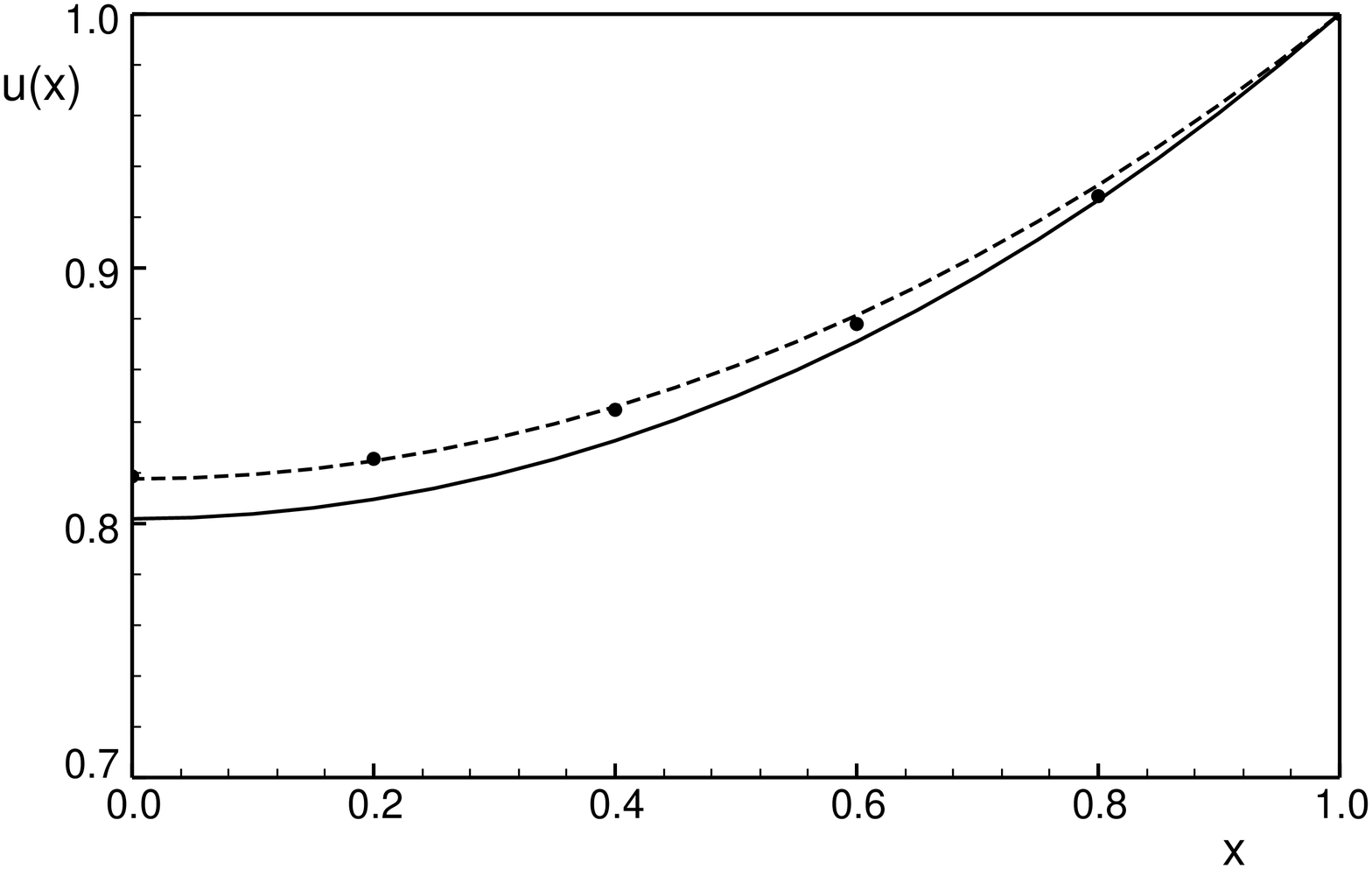}
\end{center}
\caption{Accurate values of $u(x)$ for $\epsilon=0.7$ (circles), virial
(solid line) and hypervirial (dashed line) approximate results from equation
(\ref{eq:u_app}).}
\label{fig:HT2b}
\end{figure}

Finally, in Fig. ~\ref{fig:HT2b} we compare the numerical, virial ($w=u$) and
hypervirial ($w=1$) approaches to the function $u(x)$ in a wider
scale. We conclude that the
virial theorem is not always the best choice for obtaining approximate
solutions to nonlinear problems.

\section{Conclusions}

\label{sec:conclusions}

The purpose of this paper has been the discussion of some recent results for
the nonlinear equations arising in heat transfer phenomena. The
oversimplified models considered here may probably be of no utility in
actual physical or engineering applications. Notice that the authors did not
show any sound application of those models and the only reference is a
pedagogical article cited by Rajabi et al\cite{RGT07}. However, it has not
been our purpose to discuss this issue but the validity of the methods for
obtaining exact and approximate solutions to simple nonlinear equations.

\begin{figure}[H]
\begin{center}
\includegraphics[width=9cm]{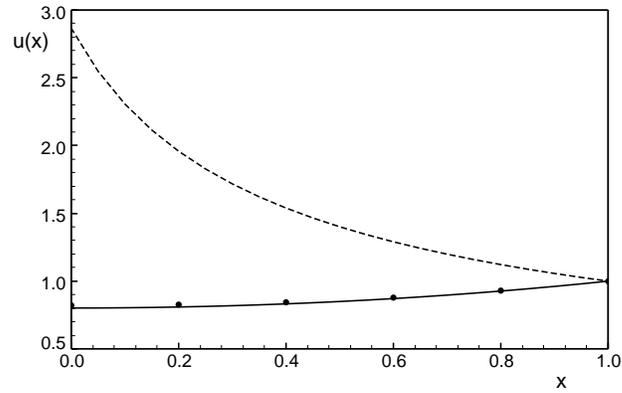}
\end{center}
\caption{Accurate values of $u(x)$ for $\epsilon=0.7$ (circles) and
approximate results from equations (\ref{eq:u_app}) (solid line) and (\ref
{eq:u_ex_2}) (dashed line).}
\label{fig:HT2}
\end{figure}

It seems that the particular application of the Lie algebraic method by
Moitsheki et al\cite{MHM09} has only produced the exact result of a trivial
equation and a wrong result for a nontrivial one. Therefore, we believe that
the authors failed to prove the utility of the technique and it is not
surprising that they concluded that their results did not agree with the HAM
ones\cite{A06c} (see Fig.~\ref{fig:HT2}).

We have also shown that under certain conditions the well known
straightforward Taylor--series method is suitable for the accurate treatment
of such nontrivial equations. It is simpler than both HAM and HPM\cite
{A06c,G06} and as accurate as the numerical integration routine built in a
computer algebra system\cite{G06}.

Finally, we have shown that the well known hypervirial theorem
may provide simple analytical expressions that are
sufficiently accurate for a successful analysis of some of those simple
models for heat transfer systems. It is surprising that our results
suggest that the virial theorem\cite {AF09a,AF09b} may not be the best
choice.

\begin{thebibliography}{9}
\bibitem{MHM09}  R. J. Moitsheki, T. Hayat, and M. Y. Malik, Comparison of
symmetry and homotopy solutions for nonlinear heat transfer systems, Appl.
Math. Comput 215:1995-2000 (2009).

\bibitem{A06c}  S. Abbasbandy, The application of homotopy analysis method
to nonlinear equations arising in heat transfer, Phys. Lett. A 360:109-113
(2006).

\bibitem{G06}  D. D. Ganji, The application of He's homotopy perturbation
method to nonlinear equations arising in heat transfer, Phys. Lett. A
355:337-341 (2006).

\bibitem{RGT07}  A. Rajabi, D. D. Ganji, and H. Taherian, Application of
homotopy perturbation method in nonlinear heat conduction and convection
equations, Phys. Lett. A 360:570-573 (2007).

\bibitem{SH08}  M. Sajid and T. Hayat, Comparison of HAM and HPM methods in
nonlinear heat conduction and convection equations, Nonlinear Analysis: Real
World Applications 9:2296-2301 (2008).

\bibitem{H04}  J-H. He, Comparison of homotopy perturbation method and
homotopy analysis method, Appl. Math. Comput. 156:527-539 (2004).

\bibitem{L05}  S. J. Liao, Comparison between the homotopy analysis method
and homotopy perturbation method, Appl. Math. Comput 169:1186-1194 (2005).

\bibitem{AF09a}  P. Amore and F. M. Fern\'{a}ndez, The virial theorem for
nonlinear problems, arXiv:0904.3858v2 [math-ph]

\bibitem{AF09b}  P. Amore and F. M. Fern\'{a}ndez, The virial theorem for
nonlinear problems, Eur. J. Phys. 30:L65-L66 (2009).
\end{thebibliography}
\end{document}